\documentclass[sn-aps]{sn-jnl}

\usepackage{changes}
\usepackage{graphicx}%
\usepackage{multirow}%
\usepackage{amsmath,amssymb,amsfonts}%
\usepackage{amsthm}%
\usepackage{mathrsfs}%
\usepackage[title]{appendix}%
\usepackage{xcolor}%
\usepackage{textcomp}%
\usepackage{manyfoot}%
\usepackage{booktabs}%
\usepackage{algorithm}%
\usepackage{algorithmicx}%
\usepackage{algpseudocode}%
\usepackage{listings}%

\theoremstyle{thmstyleone}%
\theoremstyle{thmstyletwo}%

\theoremstyle{thmstylethree}%

\begin{document}

\title[Article Title]{Influence of EOM sideband modulation noise on space-borne gravitational wave detection}


\author[1]{Mingyang Xu}

\author*[1]{Yujie Tan}\email{yjtan@hust.edu.cn}

\author[1,2]{Hanzhong Wu}

\author*[1]{Panpan Wang}\email{ppwang@hust.edu.cn}

\author[1]{Hao Yan}

\author[1]{Yurong Liang}

\author*[1]{Chenggang Shao}\email{cgshao@hust.edu.cn}

\affil[1]{MOE Key Laboratory of Fundamental Physical Quantities Measurements, Hubei Key Laboratory of Gravitation and Quantum Physics, PGMF and School of Physics, Huazhong University of Science and Technology, Wuhan 430074, China}

\affil[2]{State Key Laboratory of Applied Optics, Changchun Institute of Optics, Fine Mechanics and Physics, Chinese Academy of Sciences, Changchun 130033, China}


\abstract{Clock noise is one of the dominant noises in the space-borne gravitational wave (GW) detection. To suppress this noise, the clock noise-calibrated time-delay-interferometry (TDI) technique is proposed. In this technique, an inter-spacecraft clock tone transfer chain is necessary to obtain the comparison information of the clock noises in two spacecraft, during which an electro-optic-modulator (EOM) is critical and used to modulate the clock noise to the laser phase. Since the EOM sideband modulation process introduces modulation noise, it is significant to put forward the corresponding requirements and assess whether the commercial EOM meets. In this work, based on the typical Michelson TDI algorithm and the fundamental noise requirement of GW detectors, the analytic expression of the modulation noise requirement is strictly derived, which relax the component indicator need compared to the existing commonly used rough assessments. Furthermore, a commercial EOM (iXblue-NIR-10 GHz) is tested, and the experimental results show that it can meet the requirement of the typical GW detection mission LISA in whole scientific bandwidth by taking the optimal combination of the data stream. Even when the displacement measurement accuracy of LISA is improved to 1 pm/ $\mathrm{Hz^{1/2}}$ in the future, it still meets the demand.}




\maketitle
\section{Introduction}\label{sec1}

In 1915, Einstein proposed the general theory of relativity, giving the most elegant and precise theory of gravity to date. Its predicted GWs were directly detected by the ground-based GW detector LIGO in 2015 \cite{gw1,ligo_2}. Thus, a new era of GW astronomy was opened, and GW detection became a popular research direction for large-scale cosmic observation and basic theory test. Different GW detection methods have different sensitive frequency bands, which correspond to different wave sources and solve different scientific problems. All kinds of detection methods together help to obtain more comprehensive information about the universe. Ground-based GW detection is mainly sensitive to GWs in the frequency band of 10-$10^4$ Hz, and below 10 Hz is limited by ground vibration noise and gravity gradient noise \cite{gw1}. However, there will be more abundant GW sources below 1 Hz. Space-borne GW detection can avoid the influence of ground vibration. Meanwhile, longer arm length will be sensitive to lower frequency GW signals. For this reason, several space-borne GW detection programs have been proposed internationally, including LISA \cite{lisa}, DECIGO \cite{decigo}, Tianqin \cite{tianqin}, Taiji \cite{taiji}, etc.

Space-borne GW detection uses laser interferometry to accurately measure the phase change of the laser, traveling back and forth between the test masses placed in separated spacecraft, and then extracts the GW information from the laser interferencing scientific data streams. The sensitivity limit of the detector is determined by the instrumental noise floor, constituting by the test-mass acceleration noise and the laser shot noise. To detect a GW signal, various noises should be controlled below this floor. For a typical space-borne GW detector, laser frequency noise is the dominant noise source. In space, due to orbit motion, arm lengths cannot be equal in real time, resulting in that laser frequency noises cannot cancel each other out in the interference data stream, which greatly affects the GW detection. To solve this problem, TDI technique has been developed to construct a virtual equal-arm interferometer by time-delaying and combining the data streams, eliminating laser frequency noise in common mode\cite{tdi_1999,tdi_2002,tdi_2004,tdi1,tdi_wg_2021,tdi_wpp_2021}. 

During the digital sampling process of the heterodyne interference signal, clock noise will be also introduced. Recent years, several efforts are made to explore the clock synchronization \cite{wy,clocksy_2022,yh} and the clock-noise suppression. In this work, we focus on the latter. The characteristic Allan standard deviation of the state-of-the-art USO reads $\sigma_A\approx10^{-13}$@1 s, and this clock jitter noise is about 2$\sim$3 orders of magnitude higher than the noise floor, not meeting the requirement of the GW detection. As developing a more stable clock is difficult, TDI technique is extended to suppress clock noise. In general, there are two strategies. One is the ultrastable oscillator (USO) noise calibrated \cite{tdi_2001}, in which the laser beams are sideband modulated to construct an inter-spacecraft clock tone transfer chain, and then generate additional interspacecraft measurements about the clock noise comparison. These additional data streams can help to remove the clock noises in the scientific data streams. The other one is the optical frequency comb system connected \cite{tdi_ofc_2015}, in which the laser frequency noise is coherently linked to the clock noise, and one can modify the TDI combination to simultaneously suppress the laser and clock noises \cite{tdi_ofc_2022,tdi_ofc_2020,tdi_ofc_ole,tdi_ofc_ol}. As the onboard optical comb technology is not yet mature, this paper focuses on the former strategy.

\begin{figure}[ht!]
\centering\includegraphics[width=10cm]{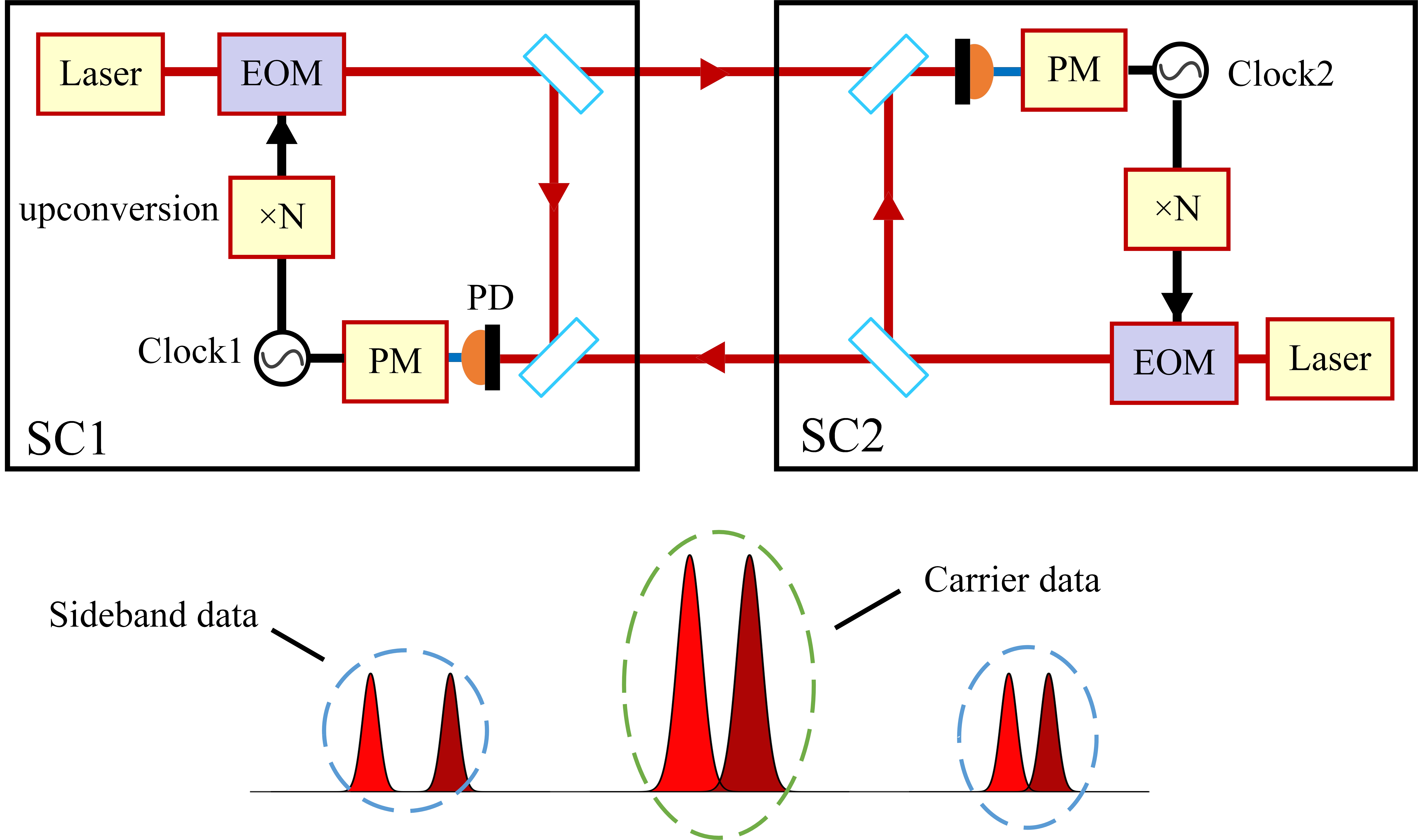}
\caption{Schematic diagram of clock sideband transmission in space-borne gravitational wave detection. The heterodyne interference signal of the two lasers is detected by the photodetector, and the phase information in the interference signal is obtained by the phasemeter. Since the phasemeter requires an external clock to trigger, the instability of the clock can also be coupled into the interferometry. In order to suppress the clock noise, the frequency of the MHz clock is multiplied to GHz by up-conversion, and then modulated to the phase of the laser by an EOM to form a sideband. The carrier and sideband transmit between the spacecrafts and interfere with the remote laser to form the carrier data stream and sideband data stream. SC: Spacecraft; PD: Photodetector; PM: Phasemeter; EOM: Electro-optic modulator.}\label{fig:1}
\end{figure}

In the USO noise-calibrated TDI technique, an EOM is a crucial experimental component, which is used to modulate the clock frequency multiplier signal to the laser phase, forming sideband data stream, as shown in Fig. \ref{fig:1}. Sideband data stream and carrier data stream are transmitted with the laser between the spacecrafts. The combination of carrier and sideband data streams is mainly dominated by clock noise, and this additional data stream can be used to eliminate clock noise in the TDI combination. Residual clock noise suppression algorithms in different TDI combinations have been reported \cite{tdi_2001,tdi_wpp_2,tdi_2018,tdi_2021,yzj}. Theoretical studies show that clock sideband TDI algorithm can suppress clock noise well below detector noise floor \cite{tdi_wpp_2}. However, in practice, the sideband modulation used to eliminate clock noise is not ideal. The homology between the frequency multiplier signal injected with the EOM and the USO \cite{cqg_2011,oe_2012}, the phase fidelity of the sideband modulation of the EOM \cite{aip_2006,apb_2010}, and the phase fidelity before and after passing through the laser amplifier will introduce additional modulation noises \cite{amplifier}. Therefore, it is important to analyze the requirement of modulation noises on the detection of GWs. Currently, the modulation noise requirement has been roughly proposed without using TDI algorithm; a typical commercial EOM has been tested, and the result showed the modulation noise in the additional data stream is large; further combined with the laser amplitude stabilization technology, it can meet the demand of GW detection \cite{apb_2010}. In this work, based on the principle of GW detection, starting from the laser interferencing data streams and TDI algorithms, we will give a more stringent and analytic expression of modulation noise requirement, and also make a related experimental test on the typical commercial EOM to see if it meets the requirements.

The paper is laid out as follows: Sec. II provides the measurement principle of typical space-borne GW detection, and the residual modulation noise after eliminating laser frequency noise and clock noise by TDI combination is derived; Sec. III shows the experiment setup for testing a commercial EOM modulation noise; Sec. IV demonstrates the experimental results. Finally, Section V is conclusion.

\section{Requirement of the sideband modulation noise}\label{sec2}
\begin{figure}[ht!]
\centering\includegraphics[width=7cm]{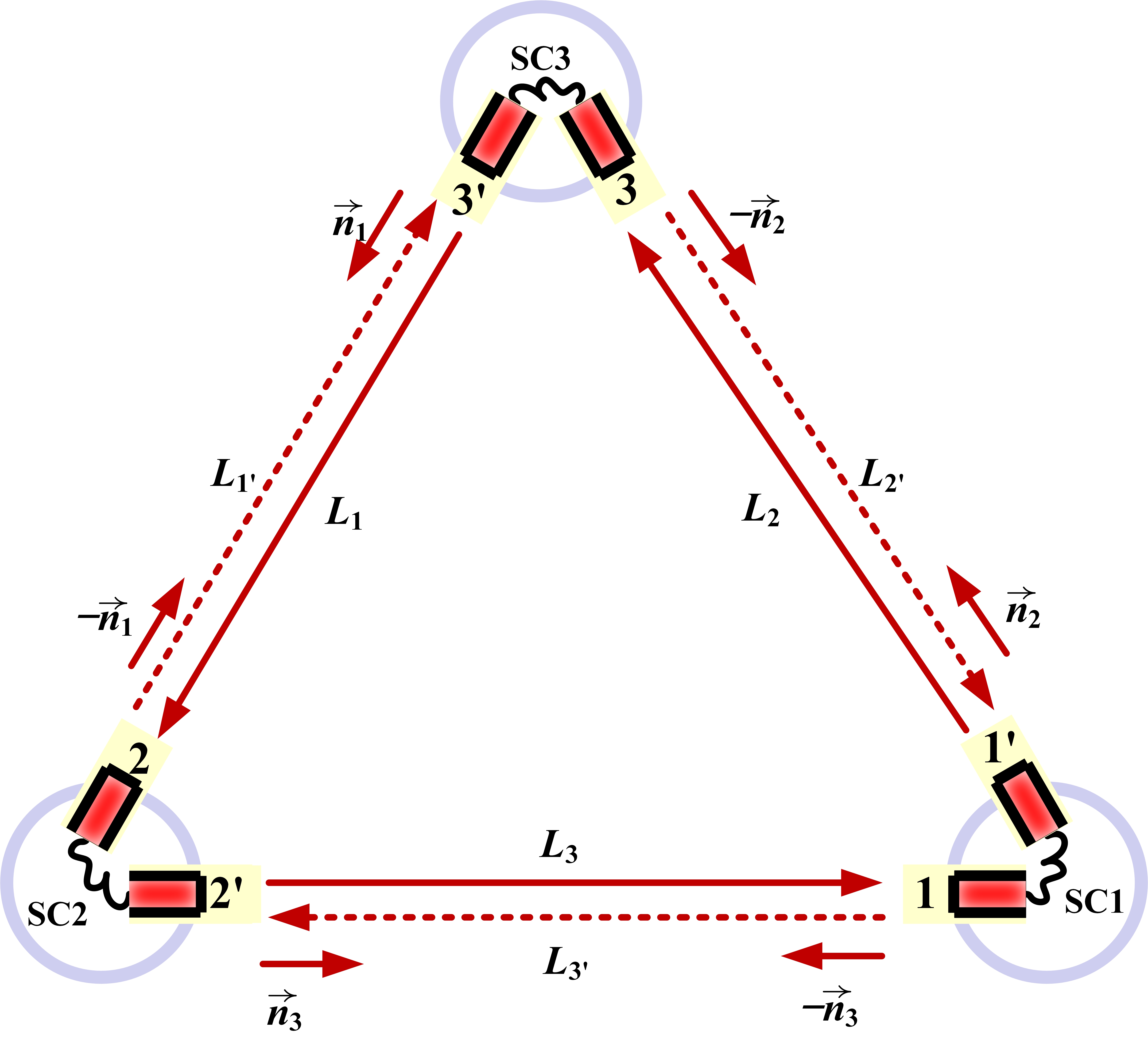}
\caption{Notation of the lasers and links in LISA.}\label{fig:2}
\end{figure}
In this section, we derive the requirements for sideband modulation noise, both notations and conventions following those defined for LISA array \cite{tdi_2003}. The structure of LISA is shown in the Fig. \ref{fig:2}. Each satellite contains two identical optical benches. It is assumed that platforms on one side are denoted as 1, 2, 3, and platforms on the other side are denoted as $1',$ $2',$ $3'$. The distance between the two satellites is denoted $L_i$ and $L_{i'}$, where $i$${=1, 2, 3,}$ denotes counterclockwise direction and $i'$${=1', 2', 3'}$ denotes clockwise direction. Each optical platform contains three-stage measurement and three data streams, which are carrier-to-carrier data stream $s_{i}^{c}$, interfered by lasers from the remote optical bench and local optical bench, which carry information about GW signals; the test mass data stream $\varepsilon_i$ , from adjacent bench to local bench, which contains information about spacecraft motion noise and test mass acceleration noise; reference data stream $\tau_i$, from adjacent optical bench to local optical bench, which only contains information about laser frequency noise, fiber noise and clock noise. Taking optical bench 1 and $1’$ as an example, these data streams can be written as \cite{tdi_wpp_2}:
\begin{align}\label{eq1}
s_{1}^{c}&=h_1+D_3p_{2'}-p_1+\left( \vec{n}_3\cdot D_3\vec{\Delta}_{2'}+\vec{n}_{3'}\cdot \vec{\Delta}_1 \right) -a_1q_1+N_{1},\nonumber\\
\varepsilon _1&=p_{1'}-p_1-2\vec{n}_{3'}\cdot \left( \vec{\delta}_1-\vec{\Delta}_1 \right) +\mu _1-b_1q_1,\nonumber\\
    \tau _1&=p_{1'}-p_1+\mu _1-b_1q_1,
    \end{align}
and
\begin{align}\label{eq2}
s_{1'}^{c}&=h_{1'}+D_{2'}p_3-p_{1'}+\left( \vec{n}_{2'}\cdot D_{2'}\vec{\Delta}_3+\vec{n}_2\cdot \vec{\Delta}_{1'} \right) -a_{1'}q_1+N_{1'},\nonumber\\
\varepsilon _{1'}&=p_1-p_{1'}-2\vec{n}_2\cdot \left( \vec{\delta}_{1'}-\vec{\Delta}_{1'} \right) +\mu _1-b_{1'}q_1,\nonumber\\
 \tau _{1'}&=p_1-p_{1'}+\mu _1-b_{1'}q_1.
\end{align}
Similarly, the data streams on other optical benches can be obtained by cyclic permutation of the indices:$1\rightarrow2\rightarrow3\rightarrow1$. Here $h_i$, $p_i$, $q_i$, $\vec{n}_i$, $\vec{\Delta}_{i}$, $\vec{\delta}_{i}$, $N_i$, and $\mu _i$ are the GW signal, laser frequency noise, clock noise, unit vectors between spacecraft, spacecraft motion noise, test mass acceleration noise, shot noise, and the fiber noise, respectively. $D_i$ and $D_{i'}$ are the time-delay operators, and for any function $f(t)$, this operator satisfies the following convention:
\begin{equation}
   D_{i'}\! D_i f(t)\!\equiv\!{D}_{i'i} f(t)\!\equiv\!{f}\!\left[\!t-\!\frac{L_{i'}(t)}{c}\!-\!\frac{L_{i}\left(t\!-\!L_{i'}(t)\right)}{c}\!\right]\!\!\approx\!{f}\!\left[\!t\!-\!\frac{L_{i'}(t)}{c}\!-\!\frac{L_{i}(t)}{c}\!\right]\!
\end{equation}
with $c$ being the speed of light. $a_i$ and $b_i$ are the coefficients corresponding to the heterodyne frequency which can be expressed as:
  \begin{align}\label{eq3}
a_i&=\frac{\nu _{\left( i+1 \right) '}-\nu _i}{f_i},\nonumber\\
a_{i'}&=\frac{\nu _{i-1}-\nu _{i'}}{f_i},\nonumber\\
b_i&=\frac{v_{i'}-\nu _i}{f_i}=-b_{i'},
\end{align}
where  $v_{i}$ represents laser center frequency and ${f_i}$ is the USO’s center frequency. By combining the carrier-to-carrier data stream, the reference data stream, and the test mass data stream, we can eliminate the spacecraft motion noise and the laser frequency noise with $i’$, and obtain:
\begin{subequations}
\begin{align}\label{eq4}
   \eta _i&\equiv s_{i}^{c}-\frac{\varepsilon _i-\tau _i}{2}-{D} _{i-1}\frac{\varepsilon _{(i+1)'}-\tau _{(i+1)'}}{2}-{D} _{i-1}\frac{\tau _{i+1}-\tau _{(i+1)'}}{2}\nonumber\\	
   &=\!h_i\!+\!{D}_{i-1}p_{i+1}\!-\!p_i\!+\!\vec{n}_{i-1}\!\cdot\!\left[ {D} _{i-1}\vec{\delta}_{(i+1)'}\!-\!\vec{\delta}_i \right]\!+\!N_i\!+\!b_{i+1}{D}_{i-1}q_{i+1}\!-\!a_iq_i,
\end{align}
\begin{align}\label{eq5}
\eta _{i^{\mathrm{'}}}&\equiv s_{i^{\mathrm{'}}}^{c}-\frac{\varepsilon _{i'}-\tau _{i'}}{2}-{D} _{(i+1)'}\frac{\varepsilon _{i-1}-\tau _{i-1}}{2}+\frac{\tau _i-\tau _{i'}}{2}\nonumber\\
&=\!h_{i'}\!+\!{D} _{(i+1)'}p_{i-1}\!-\!p_i\!+\!\vec{n}_{i+1}\!\cdot\!\left[ \vec{\delta}_{i'}\!-\!{D} _{(i+1)'}\vec{\delta}_{i-1} \right]\!+\!N_{i'}\!+\!\left( b_{i'}\!-\!a_{i'} \right) q_i.
\end{align}
\end{subequations}
Eqs. (\ref{eq4}) and (\ref{eq5}) contain three laser frequency noises $p_i$, three clock noises $q_i$, and the noise floor determined by test mass acceleration noise and shot noise. 

In order to eliminate the remaining laser frequency noise, TDI is used, which relies on properly time-shifting and linearly combining data streams in Eqs. (\ref{eq4}) and (\ref{eq5}) to construct a virtual equal arm interferometry:	
\begin{equation}\label{eq6}
    \mathrm{TDI}=\sum_{i=1}^3{\left( P_i\eta _i+P_{i'}\eta _{i'} \right)},
\end{equation}
where $P_i$ is the delay operator polynomial of different TDI combinations.
By combining Eqs. (\ref{eq4}), (\ref{eq5}) and (\ref{eq6}), we can get the residual test mass acceleration noise and shot noise as:
\begin{align}\label{eq7}
    \mathrm{TDI}^{\delta}&=\!\sum_{i=1}^3{\left\{\!-\!\left[P_i\!+\!P_{(i+1)'}{D}_{(i-1)'}\right] \vec{n}_{i-1}\!\cdot\!\vec{\delta}_i\!+\!\left[P_{i-1}{D}_{i+1}\!+\!P_{i'} \right] \vec{n}_{i+1}\!\cdot\!\vec{\delta}_{i'} \right\}},\\
    \mathrm{TDI}^{\mathrm{shot}}&=\sum_{i=1}^3{P_iN_{i}+P_{i'}N_{i'}}.
\end{align}
The residual clock noise can be also obtained as:
\begin{equation}\label{eq8}
   \mathrm{TDI}^q=-\sum_{i=1}^3{\left[ a_iP_i+a_{i'}P_{i'}-b_{i'}\left( P_{i'}-P_{i-1}{D} _{i+1} \right) \right] q_i}.
\end{equation}
Based on the assumption that the different test mass acceleration noises are independent and at the same order of magnitude, and the same assumption is made for the shot noise and clock noise, one can get the power spectral density (PSD) of test mass acceleration noise, shot noise and clock noise as:
\begin{align}
  \label{eq9}  S_{\mathrm{TDI}^{\delta}}\left( \omega \right) &=S_{\mathrm{pf}}\!\left( \omega \right) \!\sum_{i=1}^3 \left| \tilde{P}_i\left( \omega \right)\!+\!\tilde{P}_{(i+1)'}\left( \omega \right) \tilde{{D}}_{(i-1)'}\left( \omega \right) \right|^2\nonumber\\
  &+S_{\mathrm{pf}}\!\left( \omega \right) \!\sum_{i=1}^3\left| \tilde{P}_i\left( \omega \right) \tilde{{D}}_{i-1}\left( \omega \right)\!+\!\tilde{P}_{(i+1)'}\left( \omega \right) \right|^2 ,\\
 \label{eq10}   S_{\mathrm{TDI}^{\mathrm{shot}}}\left( \omega \right) &=S_{\mathrm{opt}}\left( \omega \right) \sum_{i=1}^3{\left[ \left| \tilde{P}_i\left( \omega \right) \right|^2+\left| \tilde{P}_{i'}\left( \omega \right) \right|^2 \right]},\\
  \label{eq11}  S_{\mathrm{TDl}^q}\left( \omega \right) &\!=\!S_Q\!\left( \!\omega \!\right) \!\sum_{i=1}^3{\left| a_i\tilde{P}_i\!\left(\! \omega \!\right)\! \!+\!a_{i'}\tilde{P}_{i'}\!\left( \!\omega \!\right)\! \!-\!b_{i'}\left[ \tilde{P}_{i'}\!\left(\! \omega \!\right) \!-\!\tilde{P}_{i-1}\!\left( \!\omega \!\right) \!\tilde{{D}}_{i+1}\!\left( \!\omega \!\right) \!\right] \right|^2},
\end{align}
where $S_{\mathrm{pf}}=\frac{s_{a}^{2}}{\left( 2\pi fc \right) ^2}$ , $s_a$ is the amplitude spectral density (ASD) of the test mass acceleration noise, $f$ is the Fourier frequency; $S_{\rm{opt}}=\frac{\left( 2\pi f \right) ^2s_{x}^{2}}{c^2}$  , $s_x$ is the ASD of the shot noise; $s_Q=\frac{s_q^2}{v_0^2}$, $s_q$ is the ASD of the clock noise.  $\tilde{P}_i$ represents the polynomial of the Fourier transform
of the delay operators.

With the typical TDI combinations, the residual clock noise is still about 3 orders of magnitude higher than the detector noise floor, limiting the detection of gravitational waves. To eliminate the clock noise, an inter-spacecraft clock tone modulated by an electro-optical modulator (EOM) will be used to get the information of the clock noise, which can be written as sideband-to-sideband data streams $s_{i}^{sb}$ :
\begin{equation}\label{eq12}
\begin{aligned}  
    s_{i}^{sb}=&h_i+D_{i-1}p_{{(i+1)}'}-p_i+m_{(i+1)'}D_{i-1}q_{i+1}-m_iq_i-c_iq_i+N_{i}^{sb}\\
   &+\left( \vec{n}_{i-1}\!\cdot\! D_{i-1}\vec{\Delta}_{(i+1)'}\!+\!\vec{n}_{(i-1)'}\!\cdot \!\vec{\Delta}_i \right)\!+\!m_{(i+1)'}D_{i-1}q_{i+1}^{mod}\!-\!m_iq_{i}^{mod},
    \end{aligned}
\end{equation}
where the coefficients of $m_i\approx{\rm{\frac{GHz}{MHz}}}$ are determined by the driving frequency of the EOM, $N_{i}^{sb}$ is the shot noise in sideband data streams and  $q_{i}^{mod}$ is the modulation noise. Modulation noise will enter the sideband data stream with the EOM, so the modulation noise will be introduced in the final combinations when the sideband data stream is used to eliminate clock noise. To this end, we use sideband data stream and carrier data stream to construct expressions mainly containing clock noise and modulation noise:
\begin{align}\label{eq13}
   r_i&\equiv \frac{s_{i}^{c}-s_{i}^{sb}}{m_{(i+1)'}}\approx q_i-{D} _{i-1}q_{i+1}+q_{i}^{mod}-{D} _{i-1}q_{i+1}^{mod},\nonumber\\
   r_{i'}&\equiv \frac{s_{i'}^{c}-s_{i'}^{sb}}{m_{i-1}}\approx q_i-{D} _{(i+1)'}q_{i-1}+q_{i}^{mod}-{D} _{(i+1)'}q_{i-1}^{mod}.
\end{align}

Then, we take the Michelson combination as an example to derive the residual modulation noise after eliminating the laser frequency noise and clock noise. The delay operator of the first generation of the Michelson combination is:
 \begin{align}\label{eq14}
    P_1=\left( {D} _{2'2}-1 \right), P_2=0, P_3=\left( {D} _{2'}-{D} _{33'2'} \right),\nonumber\\
    P_{1'}=\left( 1-{D} _{33'} \right), P_{2'}=\left( {D} _{2'23}-{D} _3 \right), P_{3'}=0.
    \end{align} 
   Substitute Eq. (\ref{eq14}) into Eq. (\ref{eq8}), the residual clock noise is:
   \begin{equation}\label{eq15}
   \begin{aligned}
       X_{1}^{q}=&\left[ b_{1'}\left( 1-D_{33'} \right) \left( 1-D_{22'} \right) +a_1\left( 1-D_{2'2} \right) +a_{1'}\left( D_{33'}-1 \right) \right] q_1\\
       &+\left[ a_{2'}\left( 1-{D} _{2'2} \right) {D} _3 \right] q_2-\left[ a_3\left( 1-{D} _{33'} \right) {D} _{2'} \right] q_{3'},
       \end{aligned}
   \end{equation}
   Using Eq. (\ref{eq13}), we can build auxiliary clock noise measurements:
   \begin{equation}\label{eq16}
       \begin{aligned}
           K_{X_1}=&b_{1'}\left( 1-{D} _{33'} \right) \left( r_{1'}+{D} _{2'}r_3 \right) +a_1\left( r_{1'}+D_{2'}r_3 \right) -a_{1'}\left( r_1+D_3r_{2'} \right) \\
           &+a_{2'}\left[ r_{1'}\!-\!\left( 1-D_{2'2} \right) r_1\!+\!D_{2'}r_3 \right] \!-\!a_3\left[ r_1\!-\!\left( 1\!-\!D_{33'} \right) r_{1'}\!+\!D_3r_{2'} \right] .
       \end{aligned}
   \end{equation}
   Making a combination of Eqs. (\ref{eq15}) and (\ref{eq16}), one can obtain:
   \begin{equation}
   \begin{aligned}
      X_{1}^{q}\!-\!K_{X_1}\!\approx\!&\!-\!\left[ b_{1'}\left( 1\!-\!D_{33'} \right) \left( 1\!-\!D_{22'} \right) \!+\!a_1\left( 1\!-\!D_{2'2} \right)\! +\!a_{1'}\left( D_{33'}\!-\!1 \right) \right] q_1^{mod}\\
       &-\left[ a_{2'}\left( 1-{D} _{2'2} \right) {D} _3 \right] q_2^{mod}+\left[ a_3\left( 1-{D} _{33'} \right) {D} _{2'} \right] q_{3'}^{mod}.
        \end{aligned}
   \end{equation}
   Thus, the clock noise is eliminated, while the modulation noise remains. Furthermore, the PSD of the modulation noise is:
\begin{equation}\label{eq18}
    S_{X_{1}^{mod}}\!=\!\frac{f_{i}^{2}}{\nu _{0}^{2}}4\sin ^2\!\:\!u\!\left( S_{q}^{mod} \right)^2\!\left[\! \left( \!a_1\!-\!a_{1'} \!\right) \!^2\!+\!a_{2'}^{2}\!+\!a_{3}^{2}\!+\!4b_{1'}\!\left( \!a_1\!-\!a_{1'}\!+\!b_{1'}\! \right)\! \sin ^2\!\:\!u\! \right]\!,
\end{equation}
where $S_{q}^{mod}$ is the ASD of the dimensionless
relative modulation noise. Substitute Eq. (\ref{eq14}) into Eqs. (\ref{eq9}) and (\ref{eq10}), the PSD of residual test mass acceleration noise and shot noise is:
\begin{equation}\label{eq17}
    S_{X_1}=\frac{s_{a}^{2}L^2}{u^2c^4}\left( 8\sin ^2\!\:2u+32\sin ^2\!\:u \right) +16\frac{u^2s_{x}^{2}}{L^2}\sin ^2\!\:u,
\end{equation}
where  $u=\frac{2\pi fL}{c}$ is a dimensionless quantity. For the second-generation Michelson combination $X_2$, similar to the case of the residual clock noise \cite{tdi_wpp_2}, one can find :
\begin{align}
    S_{X_2^{mod}}(\omega)&\approx4{\rm{sin}}^22uS_{X_1^{mod}}(\omega),\\
    S_{X_2}(\omega)&\approx4{\rm{sin}}^22uS_{X_1}(\omega).
\end{align}
Making $S_{X_{2}^{mod}}\left( \omega \right)\leq{S}_{X_2}\left( \omega \right) $, we can get the requirement of the sideband modulation noise as:
\begin{equation}\label{eq19}
        S_{q}^{mod}\!\leq\!\sqrt{\frac{\frac{s_{a}^{2}L^2}{u^2c^4}\left( 8\sin ^2\!\:2u+32\sin ^2\!\:u \right) +16\frac{u^2s_{x}^{2}}{L^2}\sin ^2\!\:\!u\!}{\frac{f_{i}^{2}}{\nu _{0}^{2}}4\sin ^2\!\:\!u\!\left[ \left( a_1\!-\!a_{1'} \right) ^2\!+\!a_{2'}^{2}\!+\!a_{3}^{2}\!+\!4b_{1'}\left( a_1\!-\!a_{1'}\!+\!b_{1'} \right) \sin ^2\!\:u \right]}}.
\end{equation}

In thie paper, we will take the typical parameters of LISA to analyse, the arm-length $L=2.5\times10^6$ $\rm{km}$, the ASDS of the test mass acceleration noise and shot noise are $3\times10^{-15}$ $\rm ms^{-2}/$ $\rm Hz^{1/2}$ and $10\times10^{-12}$ $\rm m/$ $\rm Hz^{1/2}$. The coefficients $a_if_i$ and $b_if_i$ are between 5 MHz and 20 MHz. Here, we take $a_1f_1=-a_{1'}f_{1}=a_{2'}f_{2}=a_{3}f_{3}=b_{1'}f_{1}$=20 MHz, and this will corresponds to the strictest requirements of modulation noise.
\begin{figure}[ht!]
\centering\includegraphics[width=10cm]{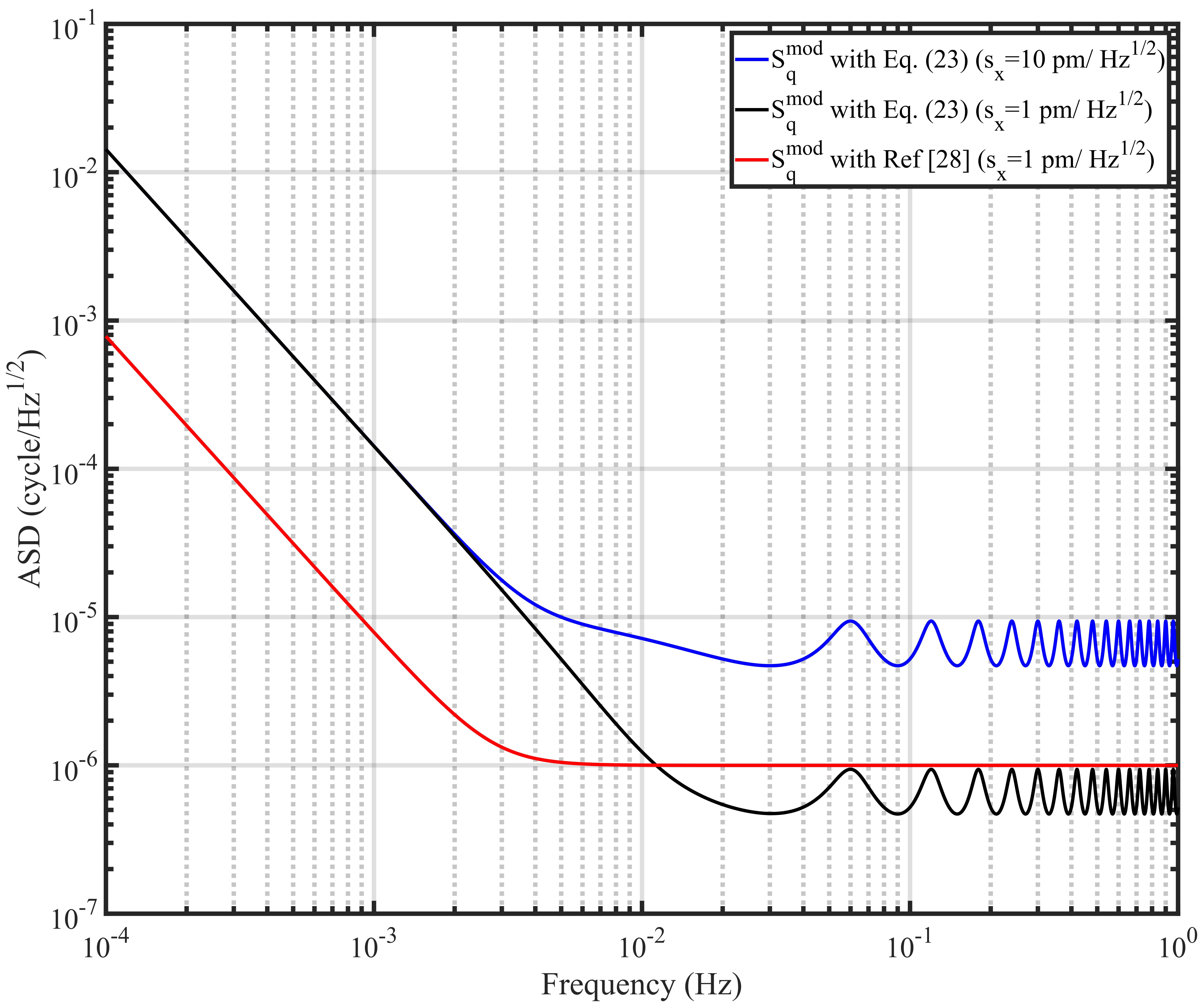}
\caption{Modulation noise requirements with typical LISA parameters.}\label{fig:3}
\end{figure}

Fig. \ref{fig:3} shows the modulation noise requirements. The dimensionless relative modulation noise requirements are multiplied by a factor of $\frac{20\rm{MHz}}{2\pi f}$ to convent into phase with the unite of cycle. The blue and black lines are respectively the modulation noise requirements according to the derived analytic expression Eq. (\ref{eq19}) with LISA parameters of $s_x=10$ $\rm pm/$ $\rm Hz^{1/2}$ and $s_x=1$ $\rm pm/$ $\rm Hz^{1/2}$. The red line is the rough assessment of the modulation noise requirement according to Ref \cite{apb_2010} with a shot noise of 1 pm/ $\rm Hz^{1/2}$. It can be clearly seen from the figure that the modulation noise requirements given by the rough evaluation are more stringent at low frequencies, below 10 mHz, than the requirements given by the analytical results. In the rough evaluation method, a shot noise level equivalent to a spacecraft displacement, such as 1 pm/ $\rm Hz^{1/2}$, is commonly allocated to the phase measurement system, while our analysis is based on the fundamental noise of gravitational wave detectors and combined with specific TDI combinations to propose indicator requirements for experimental components. The fundamental noise of gravitational wave detectors is composed of test mass noise and shot noise, and the former noise dominates at low frequencies. However, in the rough evaluation method, the indicator requirement is proposed only based on shot noise, which obviously proposes a more stringent indicator. Therefore, the analytical expression we obtained can provide a theoretical basis for the component selection for space-borne gravitational wave detectors.

\section{Experiment setup of measuring the modulation noise of EOM}
\begin{figure}[ht!]
\centering\includegraphics[width=10cm]{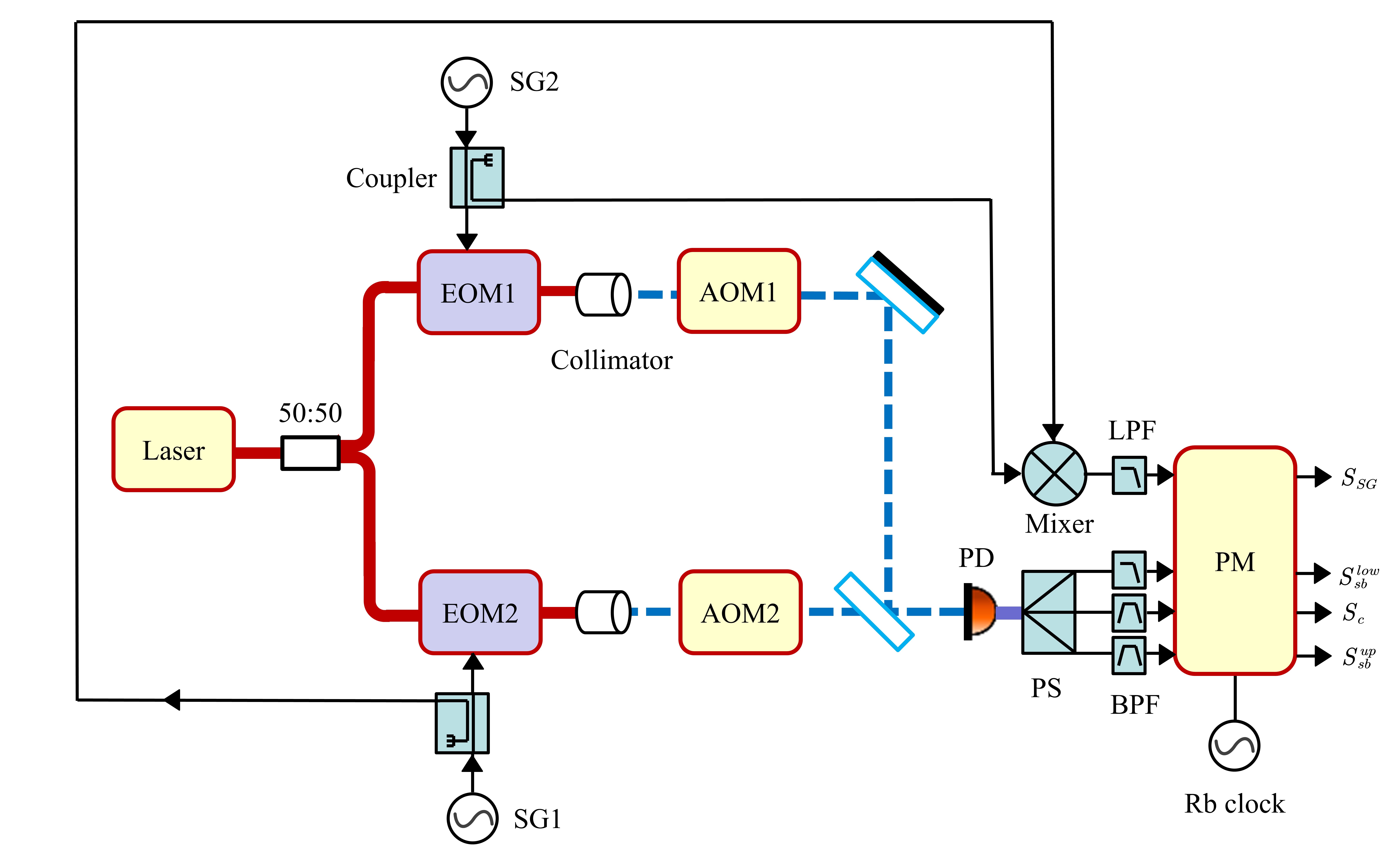}
\caption{Experiment setup for measuring the modulation noise of EOM sideband at 2.1 GHz. 50:50: fiber coupler; SG: Signal generator; LPF: Low pass filter; BPF: Band pass filter; PD: Photodetector; PS: Power splitter; PM: Phasemeter; AOM: Acousto-optic modulator.}\label{fig:4}
\end{figure}
In our experimental test, the modulation noise of the EOM is mainly measured. Other modulation noises, such as frequency divider, frequency multiplier and fiber amplifiers will be tested in the future. Fig. \ref{fig:4} shows the experimental setup used to test the EOM modulation noise. Since optical fiber is more likely to introduce ambient vibration noise and temperature fluctuation noise, we choose a spatial acousto-optic modulator (AOM) and build a spatial optical path. The whole interferometer is placed in a thermal insulation cotton box, which is used to reduce air fluctuations and temperature fluctuations. A better option would be placing the entire system in a vacuum container, but this is costly and not conducive for checking for interferometer problems. A 1064 nm laser (NKT Y10) is used as the laser signal source and two AOMs (GTCT-110 MHz) are used to generate the heterodyne interference, while the drive signals of AOM1 is 112.75 MHz and AOM2 is 107.25 MHz. We test a commercial polarization-maintaining fiber EOM (iXblue NIR-10 GHz), which has low plug loss and half wave voltage. We use the RF signal generators (Rigol DSG821) to generate signals of SG1=2.1 GHz and SG2=2.0955 GHz respectively, and through the coupler (Mini-Circuits ZX30-17-5-S+), part of which is injected into the EOM to form the upper and lower sidebands, and the other part through a mixer to get the noise of the SG itself. The two laser interferences with slightly different center frequencies and sideband can finally be received in the photodetector (Menlo FPD510-FS-NIR). We extract the upper band data stream, the carrier data stream and the lower band data stream respectively through the amplifier (Mini-Circuits ZFL-500-BNC+), the power divider (Mini-Circuits ZSC-4-2), the low-pass (DC-1.9 MHz) and band-pass filters (4.5 MHz-6.5 MHz, 9.5 MHz-11.5 MHz), and send them to the homemade FPGA phasemeter \cite{liang}. In our experiment, the SGs, the AOM drive signals and the phasemeter are referred to the common rubidium atomic clock (Stanford Research Systems FS725). 

For the experiment of Fig. \ref{fig:4}, the carrier interference data stream can be written as:
\begin{equation}\label{eq20}
    S_c=e_1\times q_1^{AOM}-e_2\times q_2^{AOM}+\delta _c,
\end{equation}
where $e_i$, ($i=1,2$) is the modulation frequency of the ${\rm{AOM}}_i$, $q_i^{AOM}$  is the noise with dimensionless relative frequency introduced by the AOM modulation driven by a SG, and  $\delta _c$ is the extra noise introduced by interferometer noise. 
The lower sideband data stream can be written as:
    \begin{align}	\label{eq21}
    S_{sb}^{low}=&e_1\times q_1^{AOM}-M_1\times q_{1}^{_{SG}}-M_1\times q_{1}^{_{EOM}}\nonumber\\	
    &-\left( e_2\!\times\! q_2^{AOM}\!-\!M_2\!\times\! q_{2}^{_{SG}}\!-\!M_2\!\times\! q_{2}^{_{EOM}} \right)\! +\!\delta _c\nonumber\\	
    =&\left( e_1\times q_1^{AOM}-e_2 \times q_2^{AOM}\right) -\left( M_1\times q_{1}^{_{SG}}-M_2\times q_{2}^{_{SG}} \right) \nonumber\\	
    &-\left( M_1\times q_{1}^{_{EOM}}-M_2\times q_{2}^{_{EOM}} \right) +\delta _c,
    \end{align}
    where $M_i$, ($i=1,2$) is the EOM modulation frequency,  $q_i^{SG}$ is the dimensionless relative frequency noise introduced by the SG, and  $q_{i}^{_{EOM}}$ is the dimensionless relative frequency noise introduced by the EOM sideband modulation. Here, we ignore the noise differences between carrier interferometers and sideband interferometers, and denote both of them as $\delta _c$. The upper sideband data stream can be written as:
    \begin{align}\label{eq22}
    S_{sb}^{up}=&e_1\times q_1^{AOM}+M_1\times q_{1}^{_{SG}}+M_1\times q_{1}^{_{EOM}}\nonumber\\	
    &-\left( e_2\times q_2^{AOM}+M_2\times q_{2}^{_{SG}}+M_2\times q_{2}^{_{EOM}} \right) +\delta _c\nonumber\\	
    =&\left( e_1\times q_1^{AOM}-e_2 \times q_2^{AOM}\right) +\left( M_1\times q_{1}^{_{SG}}-M_2\times q_{2}^{_{SG}} \right)\nonumber\\	
   & +\left( M_1\times q_{1}^{_{EOM}}-M_2\times q_{2}^{_{EOM}} \right) +\delta _c.
    \end{align}
    Thus, combining the carrier data stream and sideband data stream can eliminate the interferometer noise, and obtain the below relations:
        \begin{align}\label{eq23}
            S_{sb}^{up}-S_c&=S_c-S_{sb}^{low}=\frac{1}{2}(S_{sb}^{up}-S_{sb}^{low})\nonumber\\	
            &=\left( M_1\times q_{1}^{_{SG}}-M_2\times q_{2}^{_{SG}} \right) +\left( M_1\times q_{1}^{_{EOM}}-M_2\times q_{2}^{_{EOM}} \right).
        \end{align}
    Since the noise of SG will affect the detection, we use the mixing frequency data between the SGs to deduct the noise of the SG in the final data processing:
    \begin{equation}\label{eq25}
        \begin{aligned}
  S_{SG}&=M_1\times q_{1}^{_{SG}}-M_2\times q_{2}^{_{SG}},\\	
  \gamma _1&=S_{sb}^{up}-S_c-S_{SG}=M_1\times q_{1}^{_{EOM}}-M_2\times q_{2}^{_{EOM}},\\	
  \gamma _2&=S_c-S_{sb}^{low}-S_{SG}=M_1\times q_{1}^{_{EOM}}-M_2\times q_{2}^{_{EOM}},\\	
  \gamma _3&=S_{sb}^{up}-S_{sb}^{low}-2\times S_{SG}=2\times \left( M_1\times q_{1}^{_{EOM}}-M_2\times q_{2}^{_{EOM}} \right),
        \end{aligned}
    \end{equation}
    We assume that the noise of EOM1 and EOM2 are unrelated and at the same level, and approximately assume that the driving frequencies of EOM1 (2.1 GHz) and EOM2 (2.0955 GHz) are the same ($M$=2.1 GHz). Finally, the EOM modulation noise measured in relative frequency jitter is:
        \begin{subequations}
\begin{equation} \label{eq26a}
q_{a}^{EOM}=\frac{\gamma _1}{\sqrt{2}M},
\end{equation}
\begin{equation}\label{eq26b}
q_{b}^{EOM}=\frac{\gamma _2}{\sqrt{2}M},
\end{equation}
\begin{equation}\label{eq26c}
q_{c}^{EOM}=\frac{\gamma _3}{2\sqrt{2}M}.
\end{equation}
\end{subequations}  
where Eq. (\ref{eq26a}) is the combination of upper sideband data stream, carrier data stream and SG noise;  Eq. (\ref{eq26b}) is the combination of lower sideband data stream, carrier data stream and SG noise;  Eq. (\ref{eq26c}) is the combination of upper sideband data stream, lower sideband data stream and SG noise.
\section{Experimental Results}
\begin{figure}[ht!]
\centering\includegraphics[width=10cm]{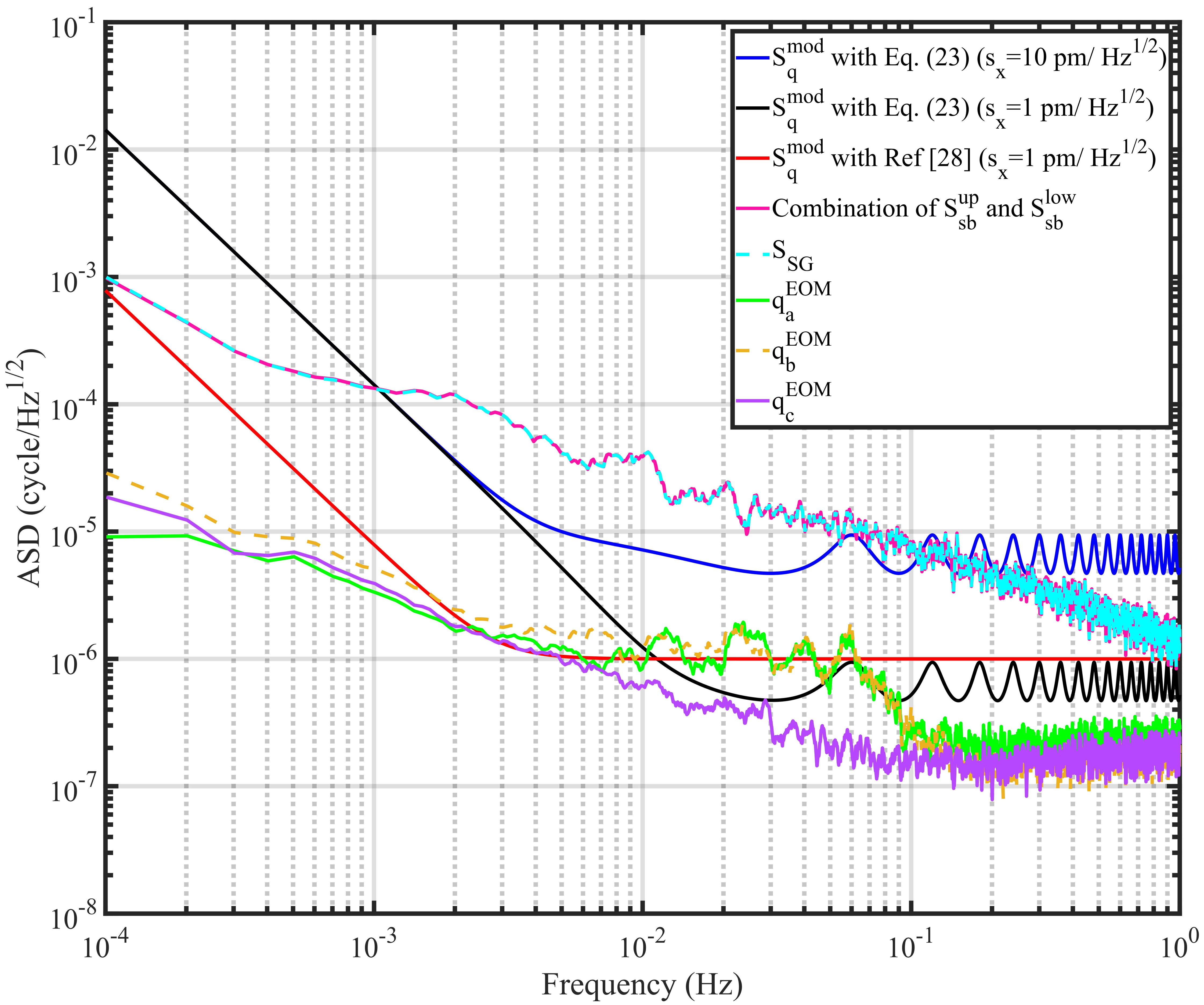}
\caption{Modulation noise of EOM experiment results.}\label{fig:5}
\end{figure}
 Fig. \ref{fig:5} shows the experimental results, in which, all the dimensionless relative frequency jitters are multiplied by a factor of $\frac{20\rm{MHz}}{2\pi f}$ to convent into phase with the unite cycle. The blue and black lines are respectively the modulation noise requirements according to the derived analytic expression Eq. (\ref{eq19}) with LISA parameters of $s_x=10$ $\rm pm/$ $\rm Hz^{1/2}$ and $s_x=1$ $\rm pm/$ $\rm Hz^{1/2}$, and the red line is the rough assessment of the modulation noise requirement according to Ref \cite{apb_2010} with a shot noise of 1 pm/ $\rm Hz^{1/2}$. The pink line represents the combination of the upper and lower sideband data, expressed as $\frac{(S_{sb}^{up}-S_{sb}^{low})}{M}\cdot\frac{20 \rm{MHz}}{2\pi f}$, which can suppress the interferometer noise. The blue dotted line is related to the noise of SG, expressed as $2\cdot\frac{{S_{SG}}}{M}\cdot\frac{20 \rm{MHz}}{2\pi f}$. Although the SGs also externally reference the same rubidium clock, the frequency multiplier inside them cannot homologate the external clock with GHz output well. The pink and blue dotted lines are trending in the same direction, indicating that they are affected by the same noise, that is, the noise of the SG, which in the experiment we have eliminated in common mode as shown in Eq. (\ref{eq25}). The green line is figured with Eq. (\ref{eq26a}) which is the combination of upper sideband data stream, carrier data stream and SG noise; the orange dotted line is figured with Eq. (\ref{eq26b}), which is the combination of lower sideband data stream, carrier data stream and SG noise; and the purple line is figured with Eq. (\ref{eq26c}) which is the combination of upper sideband data stream, lower sideband data stream and SG noise.

  In the experiment, the equivalent inter-spacecraft clock tone transfer chain can be formed by combining the upper/lower sideband and carrier data or the upper and lower sideband data. The experimental results show that the SGs multipling the clock from MHz to GHz cannot meet the requirement of LISA mission with $s_x$=10 pm/ $\rm Hz^{1/2}$ at frequencies between 1 mHz to 0.1 Hz. Fortunately, mixing data between different SGs, one can measure the contribution of SG noise to the clock tone transfer chain, and further deduct it from the chain. Based on this, we find that all the clock tone transfer chains meet the requirement of LISA mission with $s_x$=10 pm/ $\rm Hz^{1/2}$ in the whole scientific bandwidth, especially the chain formed by the upper and lower sideband data that meet the requirements of LISA mission with $s_x$=1 pm/ $\rm Hz^{1/2}$ in the whole scientific bandwidth. This may because the power jitter of the laser is coupled to the phase jitter during the EOM modulation \cite{apb_2010}, and common-mode the upper and lower sidebands can suppress the power jitter noise, since the upper and lower sidebands may have almost the same amplitude of optical power. Alternatively, one can introduce the active laser power feedback to mitigate the laser power jitter noise. Based on the above discussion and the purple line in Fig. \ref{fig:5}, the commercial EOM (iXblue NIR-10GHz) meets the requirement of LISA mission with $s_x$=1 pm/ $\rm Hz^{1/2}$, and the residual noise for the purple line may be dominated by the laser interferometer noise. 
    
 \section{Conclusion}
In space-borne GW detection, clock noise is about 2$\sim$3 orders of magnitude higher than the typical GW signal. In order to suppress the clock noise, an inter-spacecraft clock tone modulated by an EOM will be used. Theoretical studies show that clock sideband TDI algorithm can suppress clock noise well below detector noise floor. However, in practice, the sideband modulation process is not ideal, which may introduce excessive modulation noise and affect GW detection. In this work, based on the typical Michelson TDI algorithm and the noise floor of GW detectors, the analytic expression of the modulation noise requirement is strictly derived. Compared to the modulation noise requirement from the existing commonly used rough assessment, the noise requirement from the analytic expression is relaxed at the frequencies below 10 mHz. This is because the rough assessment method proposed the noise requirement only based on shot noise, while typical gravitational wave detectors are dominated by test mass noise at low frequencies. Therefore, the rough assessment method proposed a more stringent modulation noise requirement. Overall, the analytical expression we obtained can provide a theoretical basis for the component selection for space-borne gravitational wave detectors.

To evaluate whether the EOM component meet the requirements, the existing commercial EOM (iXblue NIR-10 GHz) in the laboratory has been tested. In the experiment, two commercial SGs are used to up-convert an external clock reference to GHz output. The experimental results show that the homology between the SGs and the external clock reference is not well, and the clock tone transfer chains  formed by combining the upper/lower sideband and carrier data or the upper and lower sideband data are limited by this noise. By mixing the signals from the two SGs, we construct the additional measurements for SG noises and deduct this noise from the clock tone transfer chains. Moreover, we find the differential noise between the upper and lower sideband data is lower than that between sideband data and carrier data, which may because common-mode the upper and lower sidebands can suppress the laser power jitter noise. Finally, we find the commercial EOM can meet the requirement of the typical GW detection mission LISA by taking the optimal combination of the data stream. Even when the displacement measurement accuracy of LISA is improved from 10 pm/ $\rm Hz^{1/2}$ to 1 pm/ $\rm Hz^{1/2}$ in the future, it still meets the demand. This work mainly focuses on the modulation noise introduced by the EOM, while these introduced by frequency dividers, frequency multipliers and laser amplifiers should be also analyzed to check whether the current commercial components satisfy the requirements, which will be our next research work.
\bmhead{Acknowledgments}

This work is supported by National Key Research and Development Program of China (2022YFC2204601); National Natural Science Foundation of China (11925503, 12275093 and 12175076); Natural Science Foundation of Hubei Province (2021CFB019), and State Key Laboratory of applied optics (SKLAO2022001A10).

\bmhead{Authors' contributions}

conceptualization, Mingyang Xu and Yujie Tan; methodology, Mingyang Xu and Yurong Liang; validation, Mingyang Xu, Hanzhong Wu and Hao Yan; writing—original draft preparation, Mingyang Xu; writing—review and editing, Mingyang Xu and Panpan Wang; supervision, Yujie Tan and Chenggang Shao. All authors reviewed the 
manuscript.

\bmhead{Availability of data and materials}

 Data underlying the results presented in this paper are not publicly available at this time but may be obtained from the authors upon reasonable request.

\section*{Declarations}

\bmhead{Conflict of interest} The authors declare no conflicts of interest.


\bibliography{sn-bibliography}

\end{document}